\documentclass[aps,prl,twocolumn,superscriptaddress]{revtex4-2}

\usepackage{amsmath}
\usepackage{mathtools}
\usepackage{physics}
\usepackage{siunitx}
\usepackage[pdftex]{graphicx}
\usepackage{color}

\begin{document}

\title{Search for Dark Photon Dark Matter in the Mass Range
$\mathbf{74\text{--}110}\,\si{\boldsymbol{\mu}\mathbf{eV}\boldsymbol{/c^2}}$ \\ with a Cryogenic Millimeter-Wave Receiver}

\author{S.~Kotaka}\affiliation{Department of Physics, Faculty of Science, Kyoto University, Kyoto 606-8502, Japan}
 
\author{S.~Adachi}
\email{adachi.shunsuke.5d@kyoto-u.ac.jp}
\affiliation{Hakubi Center for Advanced Research, Kyoto University, Kyoto 606-8501, Japan}
\altaffiliation[Also at ]{Department of Physics, Faculty of Science, Kyoto University, Kitashirakawa Oiwake-cho, Sakyo-ku, Kyoto 606-8502, Japan}

\author{R.~Fujinaka}
\affiliation{Department of Physics, Faculty of Science, Kyoto University, Kyoto 606-8502, Japan}

\author{S.~Honda}
\affiliation{Division of Physics, Faculty of Pure and Applied Sciences, University of Tsukuba, Ibaraki, 305-8571, Japan}
\altaffiliation[Also at ]{Tomonaga Center for the History of the Universe (TCHoU), University of Tsukuba, Japan}

\author{H.~Nakata}
\affiliation{Department of Physics, Faculty of Science, Kyoto University, Kyoto 606-8502, Japan}

\author{Y.~Seino}
\affiliation{Department of Physics, Faculty of Science, Kyoto University, Kyoto 606-8502, Japan}

\author{Y.~Sueno}
\affiliation{Department of Physics, Faculty of Science, Kyoto University, Kyoto 606-8502, Japan}

\author{T.~Sumida}
\affiliation{Department of Physics, Faculty of Science, Kyoto University, Kyoto 606-8502, Japan}

\author{J.~Suzuki}
\affiliation{Department of Physics, Faculty of Science, Kyoto University, Kyoto 606-8502, Japan}

\author{O.~Tajima}
\affiliation{Department of Physics, Faculty of Science, Kyoto University, Kyoto 606-8502, Japan}

\author{S.~Takeichi}
\affiliation{Department of Physics, Faculty of Science, Kyoto University, Kyoto 606-8502, Japan}

\collaboration{The DOSUE-RR Collaboration}

\date{\today}

\begin{abstract}
%
We search for the dark photon dark matter (DPDM) using a cryogenic millimeter-wave receiver.
DPDM has a kinetic coupling with electromagnetic fields \
with a coupling constant of $\chi$,
and is converted into ordinary photons at the surface of a metal plate. 
We search for signal of this conversion in the frequency range $18\text{--}26.5\,\si{\GHz}$, 
which corresponds to the mass range $74\text{--}110\,\si{\mu\eV}/c^2$.
We observed no significant signal excess, \
allowing us to set an upper bound of $\chi < (0.3\text{--}2.0)\times 10^{-10}$ at 95\% confidence level.
This is the most stringent constraint to date, 
and tighter than cosmological constraints. 
Improvements from previous studies are obtained by employing a cryogenic optical path and a fast spectrometer.

\end{abstract}

\maketitle

Probing the properties of cold dark matter is a crucial subject for particle physics and cosmology.
The dark matter is localized in most galaxy halos, 
but we do not understand it can interact with other standard model particles, except for via gravity. 
The dark photon is one of the dark matter candidates. 
It has a mass ($m_{\rm DP}$) and interacts with electromagnetic fields via kinetic mixing with a coupling constant of $\chi$~\cite{Wispy}.
Dark photon as dark matter (hereafter DPDM) in the mass range around $O(10\text{--}100)\,\si{\mu\eV}/c^2$ is predicted to exist in the context of high-scale inflation models~\cite{PhysRevD.93.103520}
and a part of string theories~\cite{Wispy}.
Cosmological observations give constraints in this range, with $\chi \lesssim 10^{-9}$~\cite{Wispy}.
However, the constraints set by direct searches have still not covered the wide mass range~\cite{Tomita}.

DPDM is converted to ordinary photons through the kinetic mixing at the boundary of any change in the medium~\cite{AxionLimit}, e.g., at a metal surface.
To use this characteristic, 
a methodology was suggested for searching for DPDM using an antenna~\cite{DishAntenna}.
Then, the methodology using a metal plate was established by~\cite{Tomita}.
Because the speed of DPDM ($v_{\rm DP}$) is very small compared with the speed of light ($v_{\rm DP}/c \approx 10^{-3}$), 
the direction of the conversion photons is almost perpendicular to the surface of the plate within $\sim 0.1^{\circ}$~\cite{DishDirection}. 
The conversion photons should be observed as a peak in the frequency spectrum.
The peak frequency ($\nu_0$) corresponds to the mass of the DPDM because of energy conservation, i.e., $h\nu_0 \simeq m_{\rm DP} c^2$, where $h$ is the Planck constant.
The ratio of the peak width to the peak frequency is approximately $10^{-6}$~\cite{PhysRevLett.51.1415}.

The power of the conversion photons, $P_{\mathrm{DP}}$, is given by~\cite{DishAntenna, Tomita},
\begin{eqnarray}
    P_{\text{DP}} &=& (6.4 \times 10^{-2}~{\rm aW}) \times
    \left(\frac{\chi}{10^{-10}}\right)^{2}
    \left(\frac{A_{\text{eff}}}{\SI{10}{cm^2}}\right) \nonumber \\
    && \quad \times\left(\frac{\rho}{\SI{0.39}{\GeV\per\cm^3}}\right)
    \left(\frac{\alpha}{\sqrt{2/3}}\right)^2,
    \label{eq:chi}
\end{eqnarray}
where 
$A_{\text{eff}}$ is the effective aperture area of the antenna,
$\rho = 0.39\pm0.03\,\text{GeV/cm}^{3}$ is the energy density of the dark matter in the Galactic halo~\cite{DMdensity},
and $\alpha$ is a coefficient related to the polarization of the DPDM, 
which we assume has a random orientation for, 
i.e., $\alpha = \sqrt{2/3}$~\cite{DishAntenna}. 

A previous study searched 
for DPDM in the mass range
$115.79\text{--}115.85\,\si{\mu\eV}/c^2$~\cite{Tomita} and set an upper bound of $\chi < (1.8\text{--}4.3) \times 10^{-10}$.
In this paper, we perform a similar experiment in a different mass range, $74\text{--}110\,\si{\mu\eV}/c^2$, with an improved experimental setup.

\begin{figure}[htb]
\includegraphics[width=8.6cm]{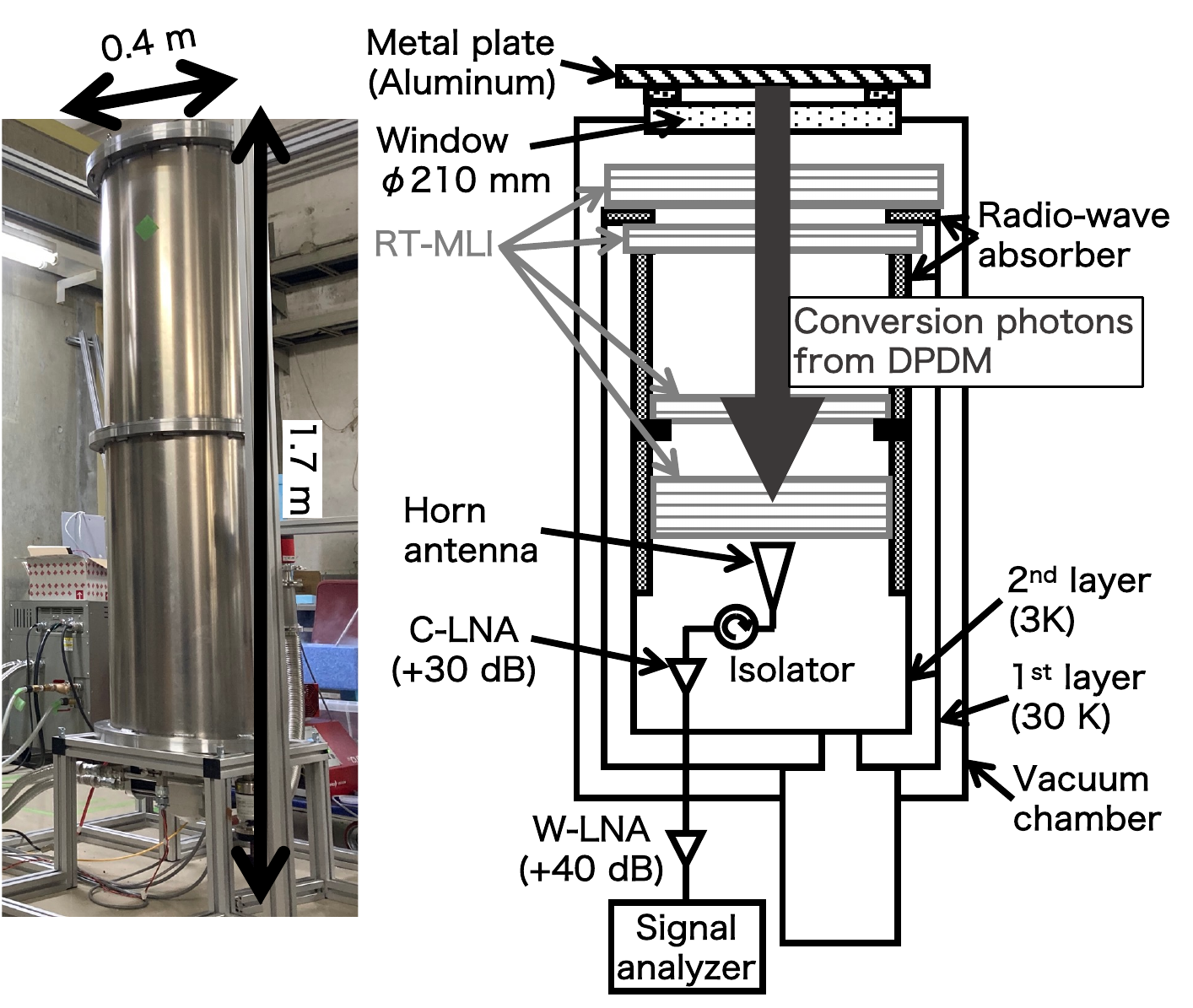}
\caption{\label{setup}
A photo and a schematic overview of the experimental setup.
A standard horn antenna (Millitech SGH-42-SC000) is used in a cryostat.
The aperture diameters of the vacuum window and the horn antenna are 210~mm and 59~mm, respectively.
We used a C-LNA (Low Noise Factory, LNF-LNC15-29B) and a W-LNA (Aldetec ALM-1826S210) to amplify detected signals. 
We obtained a spectrum using a signal analyzer, ANRITSU MS2840A.
We set the aluminum plate above the vacuum window to search for DPDM.
The distance from the plate to the antenna is 1.1~m.
We substituted calibration sources for the plate to calibrate the system.
}
\end{figure}

Our experimental setup is shown in Fig.~\ref{setup}.
We use a cryogenic millimeter-wave receiver.
DPDM is converted into ordinary photons at the lower surface of the aluminum plate. 
The conversion photons are detected by a horn antenna kept under cryogenic conditions, and the signals are amplified with both a cold low-noise-amplifier (C-LNA) and a warm one (W-LNA).
The frequency spectrum is then measured using a signal analyzer. 

To minimize the thermal noise entering the antenna, the off-axis directions from the signal are surrounded by radio absorbers kept under cryogenic conditions.
We used radio-transparent multi-layer-insulation (RT-MLI~\cite{RTMLI}) to mitigate thermal radiations from the outside. 
We succeeded in achieving 3~K and 30~K at each layer in the cryostat.
As a result, the noise temperature ($\approx 160~\mathrm{K}$) was approximately half that in the previous study whose optical path was at ambient temperature~\cite{Tomita}.
In addition, a recent spectrum analyzer allows us to increase the data-taking speed by 150 times.

A receiver gain and an offset from the receiver noise were calibrated 
by using two blackbody sources in place of the metal plate~\cite{partridge_1995}.
The blackbody sources (ECCOSORB CV3) are maintained at two temperatures: \SI{77}{\K} in the liquid nitrogen bath and room temperature as monitored ($\sim$290~K).
We measured the frequency spectra at each blackbody temperature and
we obtained the gain spectrum as a ratio 
between the difference of the measured spectra and the difference of the input powers.
The measured gains were typically $60\text{--}64\,\si{\dB}$.
The offset spectrum corresponds to the power at zero input signal from the outside,
which was also calculated using these two measurements.
The offset powers are typically $4\text{--}5\,\si{a\W}$ with a frequency bin width of \SI{2}{\kHz}.

\begin{figure}[tb]
\includegraphics[width=8cm]{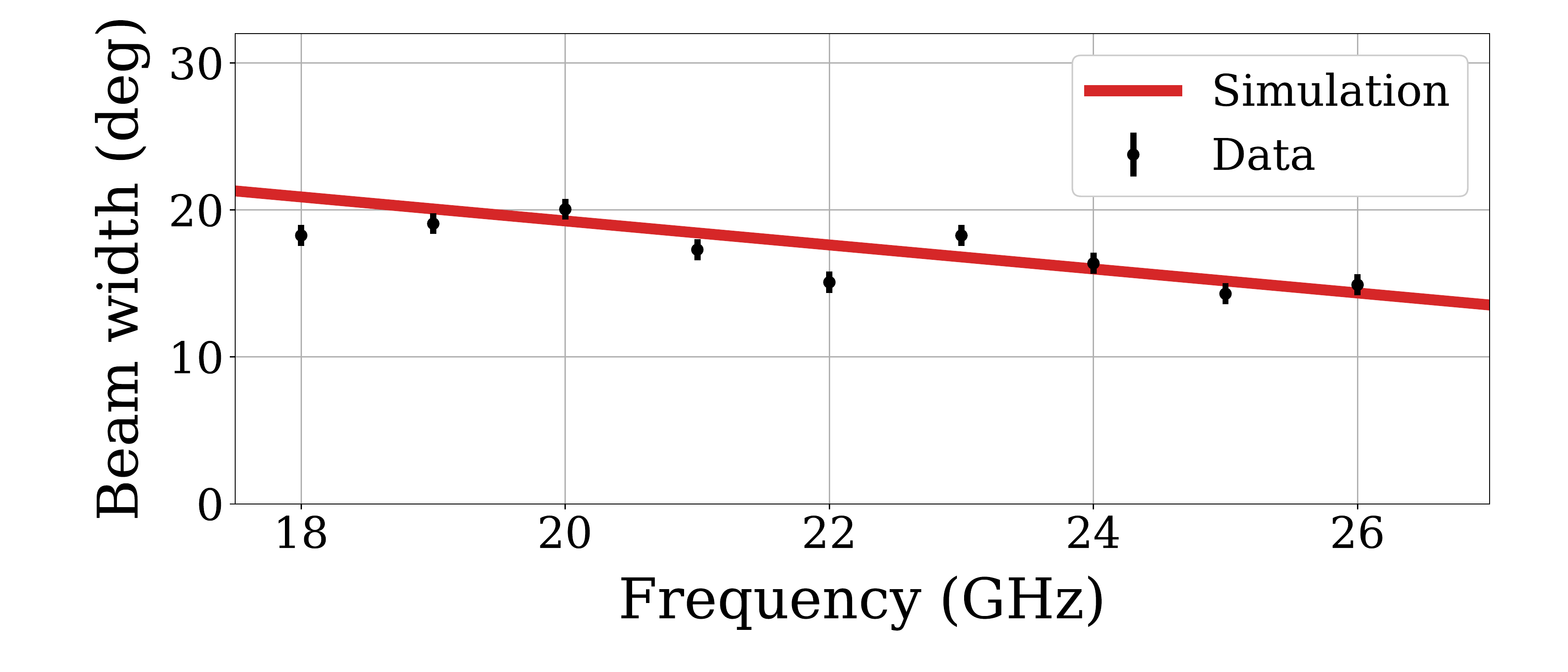}
\caption{\label{fwhm}
The beam width as a function of frequency. The error at each data point is obtained from the step of the beam scan in the calibration.
}
\end{figure}

Understanding the responsivity of the antenna as a function of angle from the line of sight (hereafter beam)
is important for calculating $A_{\rm eff}$ (the effective aperture area of the antenna)
as well as 
$\varepsilon$ 
(the fraction of the solid angle of the beam facing the blackbody sources in the calibration).  
The beam-width measurement was performed at room temperature separately from the DPDM search using an artificial source 
by combing a high-frequency signal generator (KEYSIGHT E8257D), a frequency multiplier (ERAVANT SFA-203403205-KFSF-S1), and the identical horn antenna with the receiver.
Figure~\ref{fwhm} shows the measured beam width (full width at half maximum) as a function of frequency ($\nu$).
We confirmed that the results reproduced those from a 3-dimensional electromagnetic simulation using ANSYS-HFSS. 
The remaining difference is considered to be the systematic error.
From the validated simulation, we obtained $A_{\rm eff} = \num{17.4}\pm\SI{0.3}{\cm^2}$ with negligible frequency dependence.
Similarly, we obtained $\varepsilon = \left[ 2.38 + 3.80 \times (\nu/22.0\,\si{\GHz} - 1) \pm 0.01 \right] \times 10^{-1}$ for the input power calculation from the blackbody sources. 

We took data in the frequency range $18\text{--}26.5\,\si{\GHz}$,
corresponding to $74<m_{\rm DP}<110\,\si{\mu\eV}/c^2$.
The lower and upper frequency edges were determined by the cutoff frequency of the antenna and the capability of the signal analyzer, respectively.
The signal analyzer was able to simultaneously take spectral data for a limited frequency range of \SI{2.5}{\MHz} with the resolution band width set to \SI{300}{\Hz}. 
There were 32,769 data points in each data chunk for the \SI{2.5}{\MHz} range, i.e., the frequency interval was \SI{76.3}{\Hz}.

Our initial data were taken from November 29 to December 10, 2021.
The time to accumulate the data for each chunk ($\Delta t$) was set to 2~seconds.
We shifted the center frequency by \SI{2.0}{\MHz} 
after taking 12 chunks of data for each frequency region.
In total, we took 51,000 data chunks in 4,250 frequency regions.
Regions that overlapped with the neighboring frequency region 
were used to estimate the statistical error in the analysis.
We performed the gain calibration before and after each \SI{100}{\MHz} data acquisition.
The time intervals between the calibrations were typically 40~minutes.

As described later, we obtained small values of $p_{\rm local}$ below $ 10^{-5}$ in 27 frequency regions.
Here, $p_{\rm local}$ is the local $p$-value for the zero-signal hypothesis.
For further investigation with more statistics, we took additional data with ten times longer accumulation, i.e., $\Delta t = 20$~seconds, for these regions on January 17, 2022.

We also prepared ``null samples" using the calibrated data. 
We divided the 12 data chunks into two groups for each frequency region and subtracted one from the other.
These differential samples did not contain a DPDM signal but did contain uncorrelated noise. 
There were 462 combinations for the null samples. 
Null samples have been widely used in analyses of the cosmic microwave background~\cite{QUIET2012, QUIET2013}. 
We used null samples for optimizing analysis bin width of the spectrum ($\Delta \nu = \SI{2}{\kHz}$), 
validating the statistical significance, and checking systematic uncertainties.
In the raw data, each data point has a correlation with neighbor points. We mitigated it by re-binning the data points.

\begin{figure}[tb]
\includegraphics[width=7.6cm]{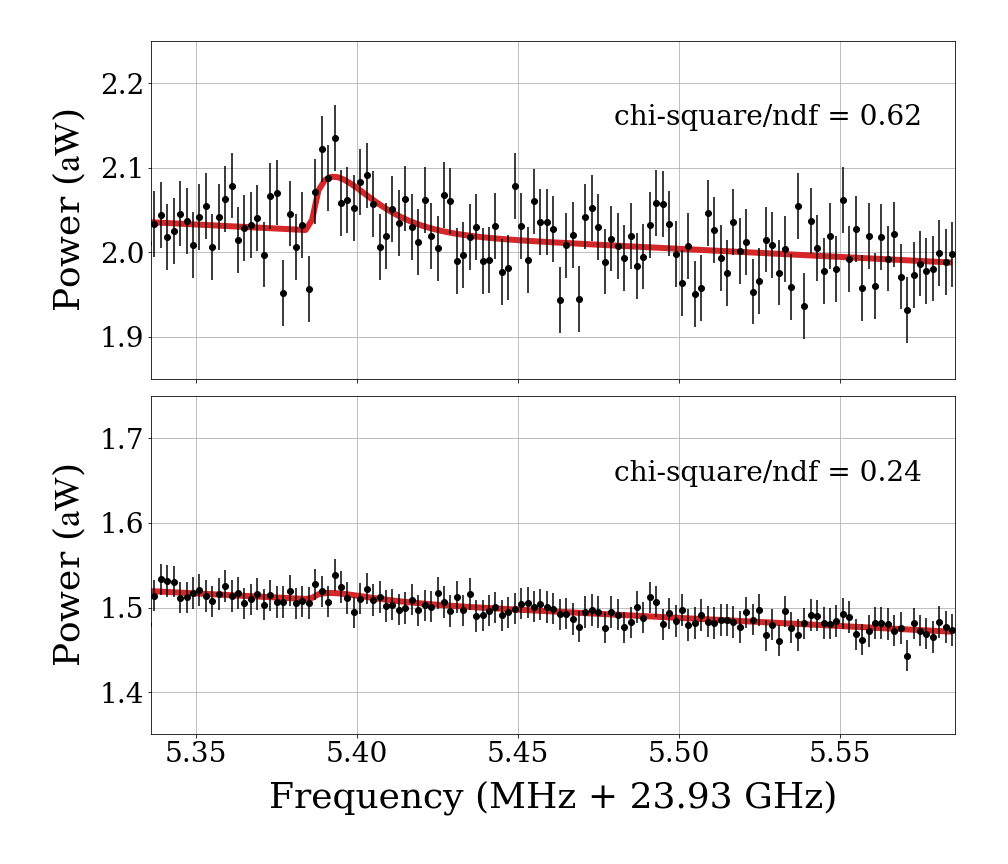}
\caption{\label{p_local_min}
Measured spectra at $\nu_0 = \SI{23.935386}{\GHz}$,
and fitted results for the signal extraction (solid line).
The upper and lower panels show the initial data
and the results with 11 times more data, respectively.
In this region, we obtained the lowest local $p$-value while the global $p$-value was not significant (see text for details).
We further confirmed that there is no signal here by taking the additional data.
}
\end{figure}

Figure~\ref{p_local_min} shows the measured spectra in one of the frequency ranges after the calibration and re-binning. 
We extracted the power of the conversion photons ($P_{\rm DP}$) by fitting for each $m_{\rm DP}$, i.e., for each $\nu_0$.
The fitting function at $\nu_0$ consisted of a signal, $f_{\rm sig} (\nu; \nu_0)$, and the background which was a one-dimensional polynomial, $f_{\rm bg} (\nu; a, b) = a (\nu - \nu_0) + b$,
\begin{equation}
 f(\nu; P_{\rm DP}, a, b) = P_{\rm DP} \times f_{\rm sig} (\nu; \nu_0) + f_{\rm bg}(\nu; \nu_0, a, b)~.
\end{equation}
$f_{\rm sig}$ is a difference in the cumulative functions, which was introduced to  account for the effect of the finite bin width,
\begin{equation}
 f_{\rm sig} (\nu; \nu_0) = F(\nu + \Delta\nu/2; \nu_0) - F(\nu - \Delta\nu/2; \nu_0)~. 
\end{equation}
The cumulative function $F$ was calculated using the following equation with the DPDM velocity and speed, $v \equiv \abs{\vb{v}} = c \sqrt{1 - (\nu_0/\nu)^2}$, 
\begin{eqnarray}
F(v) &=& \int_{0}^{v}{\text{d}v'}\int^{4\pi}{\text{d}\Omega ~g(\vb{v'};v_{\rm c}, \vb{v_{\text{E}}})~v'^2 }~,
\\
    g(\vb{v};v_{\rm c}, \vb{v_{\text{E}}}) &=&
    \frac{1}{\left(\sqrt{\pi}v_{\rm c}\right)^{3}}\exp{-\frac{|\vb{v}+\vb{v_{\text{E}}}|^2}{v_{\rm c}^2}}~,
\end{eqnarray}
where 
$g(\vb{v})$ is its velocity distribution,
$v_{\rm c}$ is the circular rotational speed of the Galaxy,
and $\vb{v_{\text{E}}}$ is the velocity of the Earth in the frame of the Galaxy.
We assumed a Maxwell-Boltzmann distribution for $g(\vb{v})$~\cite{DMVelocityDistribution}.
We also assumed $\abs{\vb{v}_{\rm E}} = v_{\rm c} = 220~{\rm km/s}$, as in many dark matter searches~\cite{DMVelocity,LUX,Xenon1T, PandaX,DarkSide}.
The width of $f_{\rm sig}$ was approximately \SI{20}{\kHz}, following from the above.
Further details on the signal distribution were described in~\cite{Tomita}.

We varied the peak frequency $\nu_0$ from \SI{18.0}{\GHz} to \SI{26.5}{\GHz} 
in a small steps of
$\Delta \nu = \SI{2}{\kHz}~(\ll \text{signal width})$,
and performed a fit with floating $P_\mathrm{DP}$, $a$, and $b$, in the frequency range
from $\nu_0 - \SI{50}{\kHz}$ to $\nu_0 + \SI{200}{\kHz}$, as shown in Fig.~\ref{p_local_min}.
We calculated the standard deviations of the data for the \SI{250}{\kHz} regions below and above the fit range, and their average was assigned to be the error associated with each data point in the fit.

\begin{figure}[tb]
\includegraphics[width=8.6cm]{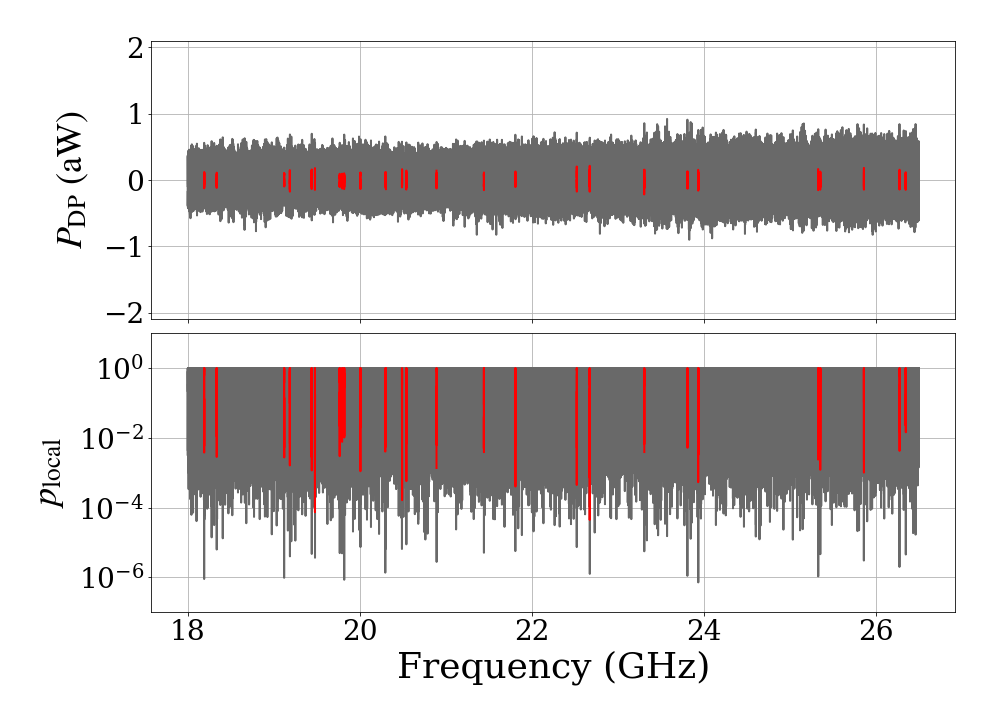}
\caption{\label{p_local}
The extracted signal powers (top) and their local $p$-values (bottom) at each frequency.
The results from the initial data are shown in gray and those obtained with additional data are shown in red.
}
\end{figure}

Before the above signal extraction, we performed fits to the null samples in the same manner.
For each null sample at each frequency, we obtained $P_{\rm DP}$, the error $\sigma$, and their ratio $x = P_{\rm DP}/\sigma$.
We obtained a zero-consistent mean value for the distribution of $x$, as expected,
and we did not observe any frequency dependence.
Thus, the normalized distribution of the ratios, $\mathcal{P}(x)$, was used to estimate the local $p$-values for the zero-signal hypothesis~\cite{pdg},
\begin{equation}\label{eq:local-p}
 p_{\text{local}} = \int_{x}^{\infty}{\mathcal{P}(x')}\text{d}x'~.
\end{equation}

Figure~\ref{p_local} shows the extracted $P_{\rm DP}$ at each frequency and their $p_{\rm local}$. 
The minimum value was $p_{\rm local}^{\rm min} = 7.1 \times 10^{-7}$ at $\nu_0 = \SI{23.935386}{\GHz}$.
Adopting a methodology similar to those in previous studies~\cite{Tomita, PhysRevD.97.123006}, we accounted for the look-elsewhere-effect. 
We determined the number of independent frequency windows ($1.6 \times 10^6$) using the null samples.
The probability of exceeding $p_{\rm local}^{\rm min}$ at any frequency was estimated as, 
\begin{equation}
    p_{\rm global} = 1 - (1- p_{\rm local}^{\rm min})^{1.6 \times 10^6} = 0.68~.
\end{equation}
We did not observe any significant excess of the DPDM signal from zero.

In the initial data set, we found 27 frequency regions with $p_{\rm local} < 10^{-5}$. 
To obtain a more robust conclusion, we additionally took data for these regions giving 11 times more statistics.
As shown in the bottom plot in Fig.~\ref{p_local_min}, 
we obtained zero-consistent results at the frequency with the minimum $p$-value in the initial data set.
In the other regions, we obtained less significant $p$-values ($p_{\rm local} > 10^{-5}$) as shown in Fig.~\ref{p_local}.
We conclude that there is no significant signal from DPDM in this search.

The systematic uncertainties associated with the coupling constant $\chi$ are summarized in Table~\ref{tab:syst}.
The uncertainty from $A_{\rm eff}$
was estimated from the difference in the beam width between the calibration and simulation.
The uncertainty assigned to the gain conservatively includes the maximum variation across all the gain calibration intervals during the experiment (3.4\%), 
the difference in $\varepsilon$ between the calibration and the simulation (1.9\%),
and uncertainties of the source temperature and emissivity ($0.7\%$).
A possible fitting bias due to the frequency binning was estimated using the simulation.
For the instrumental alignment, 
the tilt of the plate to the antenna was at most \ang{0.05}.
The alignment contributes only a small systematic error to $\chi$ because the beam width is large as shown in Fig.~\ref{fwhm}.
The uncertainty related to the direction of the conversion photons was similarly obtained.
For the dark matter density, we used the uncertainty described in \cite{DMdensity}.

\begin{table}[tb]
\caption{\label{tab:syst}
Systematic uncertainties associated with the coupling constant $\chi$.
}
\begin{ruledtabular}
\begin{tabular}{lr}
    Source                  & ($\%$) \\
    \colrule
    \textrm{Effective aperture area ($A_{\rm eff}$)}    & 4.2 \\
    \textrm{Gain}                                       & 4.0 \\
    \textrm{Frequency bin}                              & 0.6 \\
    \textrm{Alignment of instruments}                   & $<$ 0.1 \\
    \textrm{Direction of conversion photons}            & $<$ 0.1 \\
    \textrm{Dark matter density ($\rho$)}               & 3.9 \\
    \colrule
    \textrm{Total}                                      & 7.0 \\  
\end{tabular}
\end{ruledtabular}
\end{table}

\begin{figure}[tb]
\includegraphics[width=8.6cm]{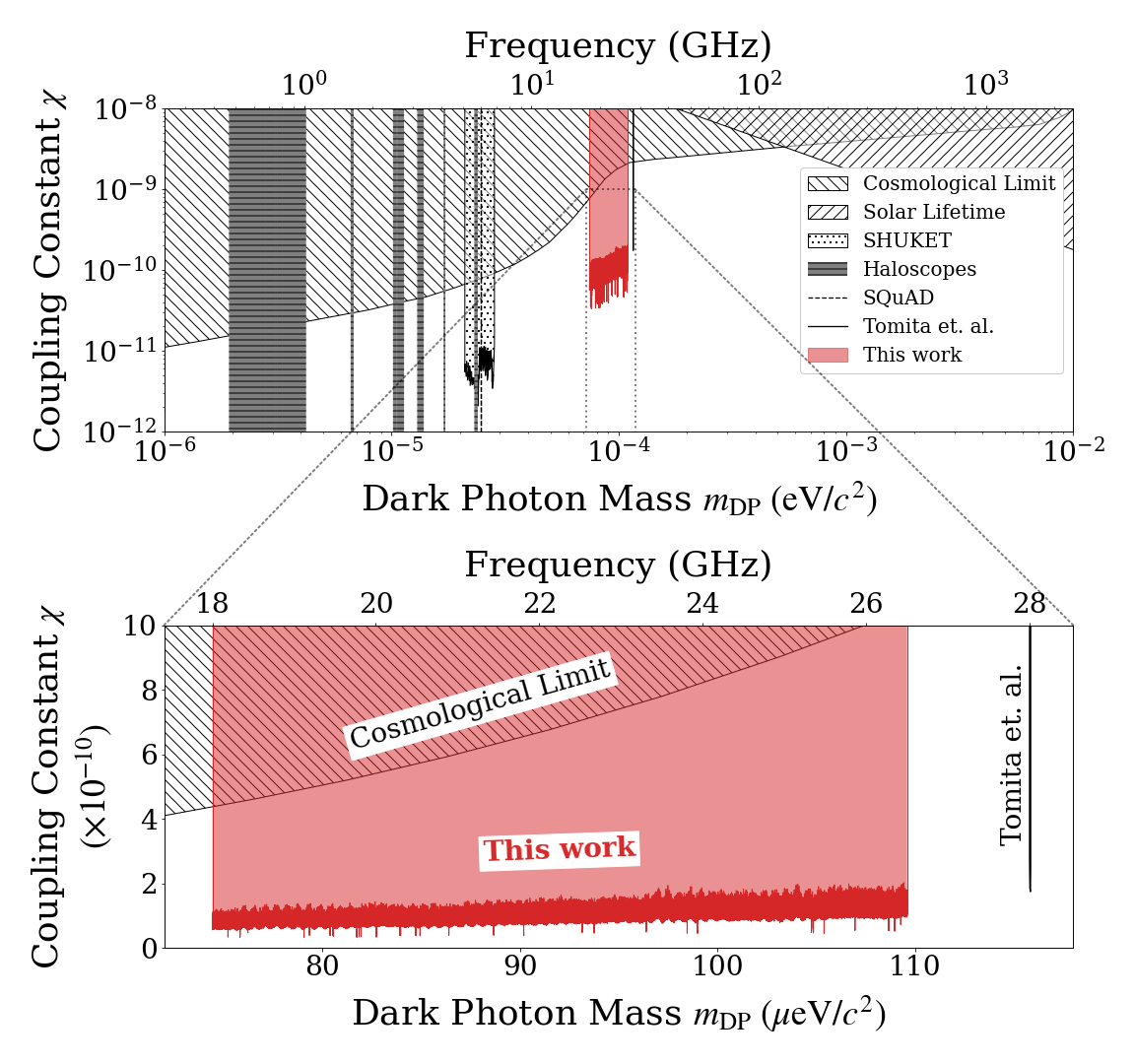} 
\caption{\label{chi_limit}
Constraints for $\chi$ at 95\% confidence level as a function of $m_\mathrm{DP}$.
Results of previous research are obtained from~\cite{Tomita, SHUKET, SQuAD, AxionLimit}.
}
\end{figure}

The upper bounds on $P_{\rm DP}$ at 95\% confidence level for each frequency were also calculated using the $\mathcal{P}(x)$, 
\begin{equation}
    {\rm max}(0, P_{\rm DP}) + 1.71 \sigma~.
\end{equation}
Here, the value of 1.71 is slightly larger than that of the normal Gaussian (1.65)~\cite{pdg}. 
This is due to the distribution tail in $\mathcal{P}(x)$.
The upper limits on $P_{\rm DP}$ were converted into the upper limits on $\chi$ using Eq.~(\ref{eq:chi}).
The systematic uncertainty was also considered in this process.
As shown in Fig.~\ref{chi_limit}, we obtained limits for $\chi < (0.3 \text{--} 2.0) \times 10^{-10}$ at a $95\%$ confidence level in the mass range $74\text{--}110\,\si{\mu\eV}/c^2$.
This is the most stringent constraint to date, and tighter than that given by cosmological observations.

In summary, we performed a search for DPDM using a cryogenic receiver in the millimeter-wave range, $18\text{--}26.5\,\si{\GHz}$, which corresponds to a mass range $74\text{--}110\,\si{\mu\eV}/c^2$.
We optimized the analysis procedure using null samples and calculated the statistical significance using them.
We found no signal and set an upper limit of $\chi < (0.3\text{--}2.0) \times 10^{-10}$ at $95\%$ confidence level.
This is the first exploration of a mass range that had not yet been explored by any direct search.
The constraint achieved is tighter than that from the cosmological observations.
The explored mass range is 600 times larger than that in the previous study with an improved setup.

This work was supported by JSPS KAKENHI under grand numbers 
20K14486,
20K20427,
21H01093,
and
21H05460,
and was also supported by grant aid from the Murata Foundation and the Sumitomo Foundation.
SA and TS acknowledge the Hakubi Project and the SPIRITS Program of Kyoto University, respectively.
We thank Edanz (\url{https://jp.edanz.com/ac}) for editing a draft of this manuscript.


\providecommand{\noopsort}[1]{}\providecommand{\singleletter}[1]{#1}%

\end{document}